\documentclass{emulateapj}
\pdfoutput=1
\usepackage[colorlinks=true,breaklinks=true,citecolor=blue,linkcolor=blue,urlcolor=blue]{hyperref}

\begin{document}

\shorttitle{Exo-Trojans}
\shortauthors{Michael Hippke \& Daniel Angerhausen}
\title{A statistical search for a population of Exo-Trojans in the Kepler dataset}

\author{Michael Hippke}
\email{hippke@ifda.eu}
\affil{Luiter Stra{\ss}e 21b, 47506 Neukirchen-Vluyn, Germany}

\author{Daniel Angerhausen}
\email{daniel.angerhausen@nasa.gov}
\affil{NASA Postdoctoral Program Fellow, NASA Goddard Space Flight Center, Greenbelt, MD 20771, USA}

\begin{abstract}
Trojans are small bodies in planetary Lagrangian points. In our solar system, Jupiter has the largest number of such companions. Their existence is assumed for exoplanetary systems as well, but none has been found so far. We present an analysis by super-stacking $\sim4\times10^4$ \textit{Kepler} planets with a total of $\sim9\times10^5$ transits, searching for an average trojan transit dip. Our result gives an upper limit to the average Trojan transiting area (per planet) corresponding to one body of radius $<460$km at $2\sigma$ confidence. We find a significant Trojan-like signal in a sub-sample for planets with more (or larger) Trojans for periods $>$60~days. Our tentative results can and should be checked with improved data from future missions like \textit{PLATO~2.0}, and can guide planetary formation theories.
\end{abstract}

\section{Introduction}
\label{sec:introduction}
\begin{figure}
\includegraphics[width=\linewidth]{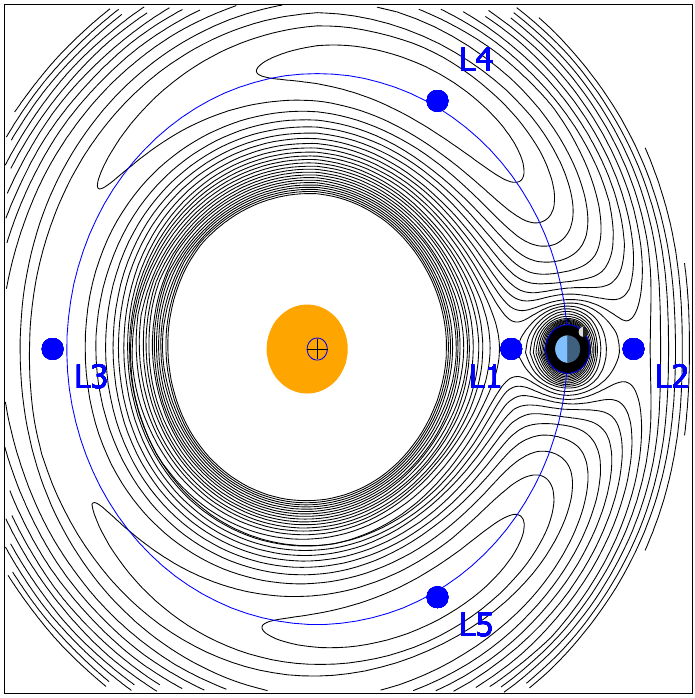}
\caption{\label{fig:scheme}Lagrangian points L4 and L5 are at 60$^\circ$ from the planet.\\}
\end{figure}

Back in 1771, the mathematician Lagrange found a solution of the three-body-problem for a primary planet and an asteroid of small mass. When the bodies are in the same plane in circular orbits of the same period, the stable locations for the asteroid are 60$^\circ$ from the planet \citep{Lagrange1772}. As no such asteroids where known at the time, the problem was considered to be only of mathematical interest. Today, we refer to these points as the (stable) Lagrangian points L4 and L5 (Figure~\ref{fig:scheme}, based on \citet{Cornish1998}). 

More than a century later, Max \citet{Wolf1906} of the University Observatory Heidelberg discovered a new ``planet'' 55$^\circ$ east of Jupiter and immediately noted its strange orbit: ``the small change in R.A. is remarkable''\footnote{In original German language: ``Bemerkenswerst ist (...) die kleine RA.-Bewegung von TG''}. More such bodies were found in the same year, and it was realized quickly that these are trapped in Jupiter's Lagrangian points. To distinguish them from the main belt asteroids, which usually receive female names, it was decided to name them after Greek heroes of the Trojan war. Wolf's ``planet'' is today known as (588)~Achilles, and is in the L4 group.

Asteroids that are trapped in L4 or L5 orbit around their point of equilibrium in a tadpole, or horseshoe orbit \citep{Marzari2002}. Today, $>6,000$ Jupiter Trojans are known\footnote{IAU Minor Planet Center, \url{http://www.minorplanetcenter.net/iau/lists/JupiterTrojans.html}, list retrieved on 26-Apr 2015}, as well as a few Neptune-, Mars- and Earth Trojans. The largest known Trojans have sizes $>100$km in radius \citep{Fernandez2003}, and it is believed that the total number of L4 Jupiter Trojans, with radii $>1$km, is $\sim6\times10^5$ \citep{Yoshida2005}. If L5 contains an equal amount of debris, then the total transiting area equivalent of small Jupiter Trojans is corresponding to one body of radius $\sim$600km. The 32 largest objects \citep{Fernandez2003} account for an additional radius equivalent of $\sim$300km. 

The Lagrangian points are stable over Gyr timescales, as long as the planet is $<4$\% of the system mass \citep{Murray1999}. Most of the system mass is usually concentrated in the host star, e.g. 4\% of $M_{\odot}$ is 40$M_{Jup}$, so that this limit is usually met. Consequently, we might assume that other planetary systems also posses Trojan bodies; this is also expected from formation mechanisms in protoplanetary accretion discs
\citep{Laughlin2002}. As the properties of extrasolar systems are diverse, we can ask the question of how large these bodies can be, and how many there are. There is nothing that physically prevents them from occurring in larger numbers (and/or larger sizes) than in our own system. Hypothetical exo-Trojans have been shown to be stable for up to Jupiter mass in the most extreme cases \citep{Erdi2007}, assuming low eccentricity \citep{Dvorak2004}.

Searching for Trojans in time-series photometry is difficult, as these bodies librate around their equilibrium points to a substantial degree. This produces large transit timing variations, so that they are missed in standard planet-search algorithms. Also, the mean inclination of Jupiter Trojans is 10$^\circ$ \citep{Yoshida2005} to 14$^\circ$ \citep{Jewitt2000}. If this is typical for exo-Trojans, only part of the swarm would go into transit. 

Data from the \textit{Kepler} space telescope have been searched for individual Trojans, with a null result and sensitivity down to $\sim1R_{\oplus}$ \citep{Janson2013}. Another search was carried out with data from the \textit{MOST} satellite for the transiting Hot Jupiter HD~209458b, also with a null result and an upper limit of $\sim1$ lunar mass of asteroids \citep{Moldovan2010}.

Although interesting, we do not repeat these searches for individual Trojans here, but ask the question of the \textit{average} Trojan effect in all \textit{Kepler} data. Millions of small Trojans might show up, on average, when stacking $\sim4\times10^4$ planets with a total of $\sim9\times10^5$ transits, as is the case for exo-moons \citep{Hippke2015}.

\section{Method}
\label{sec:method}
We employed the largest database available: High precision time-series photometry from the \textit{Kepler} spacecraft, covering 4 years of observations \citep{Caldwell2010}. 

\subsection{Data selection}
Based on a list of all validated transiting \textit{Kepler} planets (821) and un-validated planet candidates (3,359)  \citep{Wright2011}\footnote{www.http://exoplanets.org, list retrieved on 18-Nov 2014}, we downloaded their \textit{Kepler} long-cadence (LC, 30min) datasets. We used the same dataset as published by the Transiting Planet Search (TPS) pipeline, which relies on a systematic error-corrected flux time series from a ``wavelet-based, adaptive matched filter that characterizes the power spectral density (PSD) of the background process yielding the observed light curve and uses this time-variable PSD estimate to realize a pre-whitening filter and whiten the light curve''\citep{Borucki2011}. This dataset was used for most planet validations (e.g., \citet{Lissauer2012, Rowe2014}). We have downloaded these data from the NASA Exoplanet Archive\footnote{\url{http://exoplanetarchive.ipac.caltech.edu/docs/API_tce_columns.html}, retrieved on 21-Apr 2015} and applied no further corrections or detrending. It must be assumed that there are unidentified transits, and stellar trends, in these data, but we can also assume that these are distributed randomly over phase time, so that no systematic effect should affect the precise locations of the L4 and L5 phase time.

\subsection{Data processing}
Each planet has its own light curve in this dataset, which comes with companion transits removed (in multiple systems). We phase-folded every lightcurve with its published period. Afterwards, we re-normalized the data for each curve, while masking the times around planetary primary and secondary transit. Then, we re-binned each phase-folded lightcurve in 1,000 bins. Depending on the period, this is equivalent to a time of 1min (for the shortest period) to 18hrs (for a 750-day period). For the median period of 13 days, the bin length is 20min. As the average transit duration is a few hours, smearing occurs only for the few very long period planets.

\subsection{The super-stack}
From the sample of 3739 useful phase-folded lightcurves in 1,000 bins, we created a super-stack by co-adding these, and taking the median of each bin. This method was also used by \citet{Sheets2014} for the detection of average secondary eclipses, and by \citet{Hippke2015} for the search of an average exo-moon effect. In contrast to these studies, we did not stretch to the expected transit duration, because Trojans are expected to be in orbits around the Lagrange-points, and not stationary in phase time.

The resulting data were strongly dominated by outliers. This is caused by many factors contributing to different noise levels: The brightness of the host star, stellar variability, instrumental differences, and others. We decided on two filters to remove outliers, namely the stellar brightness (we kept stars brighter than 15mag in \textit{J} as measured by 2MASS), and the scatter per star (we kept the better half).

It is interesting to mention a (slight) selection effect from this filter choice: When rejecting dimmer and/or noisier stars, the average stellar radius changes. Smaller stars (e.g. M-dwarfs) exhibit more stellar noise \citep{Basri2013} and are usually less luminous. Consequently, our sample is shifted towards larger stellar radii. While the total \textit{Kepler}-planet sample has an average stellar radius of 1.14$R_{\odot}$, our post-filter sample has an average of 1.17$R_{\odot}$.

\section{Results}
\label{sec:results}
\begin{figure}
\includegraphics[width=\linewidth]{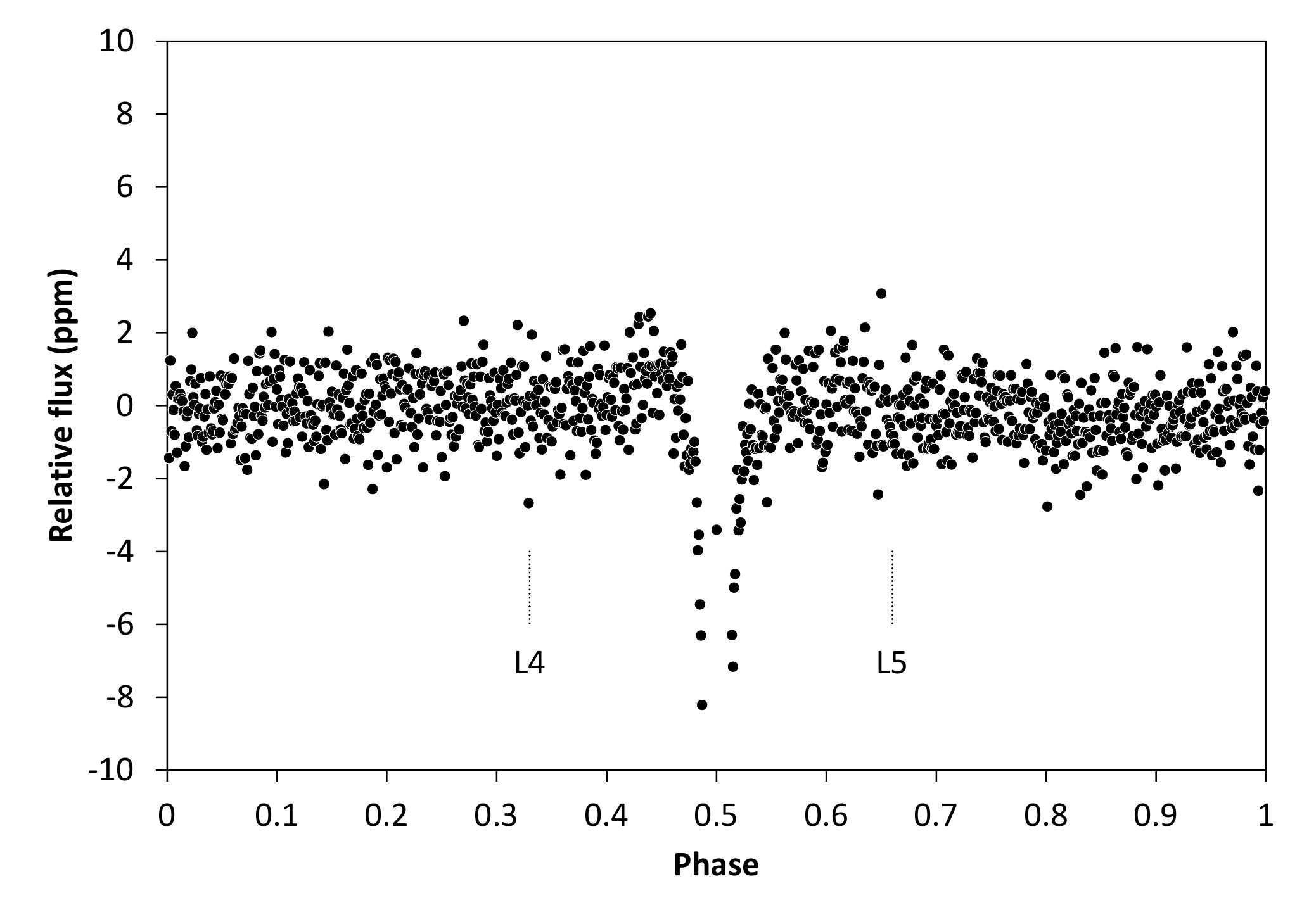}
\caption{\label{fig:fold}The initial superstack shows no significant dips at the Lagrangian points.\\}
\end{figure}

The initial post-filtered superstack does not show any significant Trojan dips, as shown in Figure~\ref{fig:fold}. When taking the average flux in a bin of 0.03 width in phase space, we obtain $+0.06\pm0.23$ppm for L4, and $-0.10\pm0.23$ppm for L5. For the average stellar radius of 1.17$R_{\odot}$, we can set an upper limit for the average Trojan area (per planet) of 460km at $2\sigma$ confidence. This applies to the full (filtered) \textit{Kepler} sample.

\begin{figure*}
\includegraphics[width=0.5\linewidth]{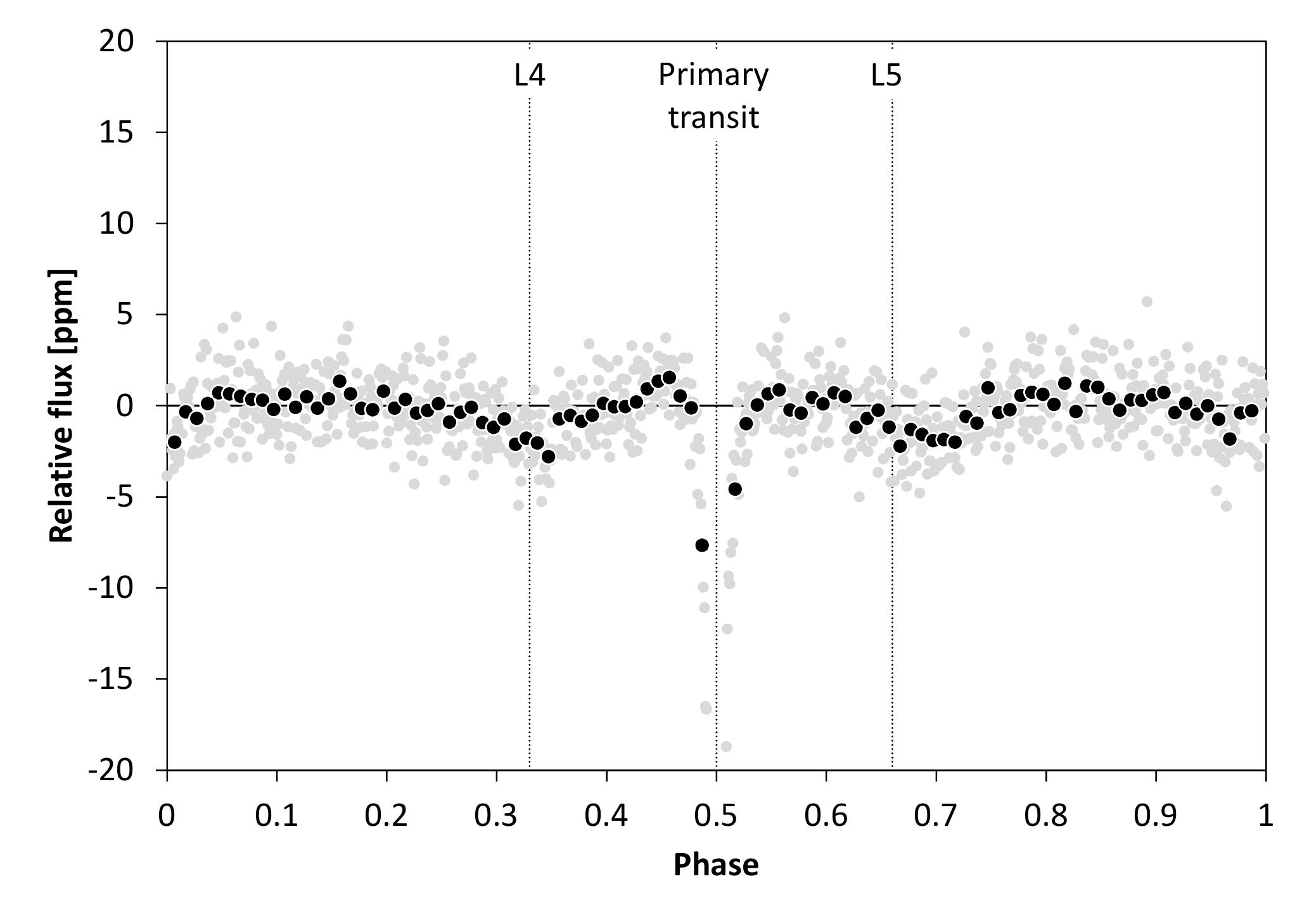}
\includegraphics[width=0.5\linewidth]{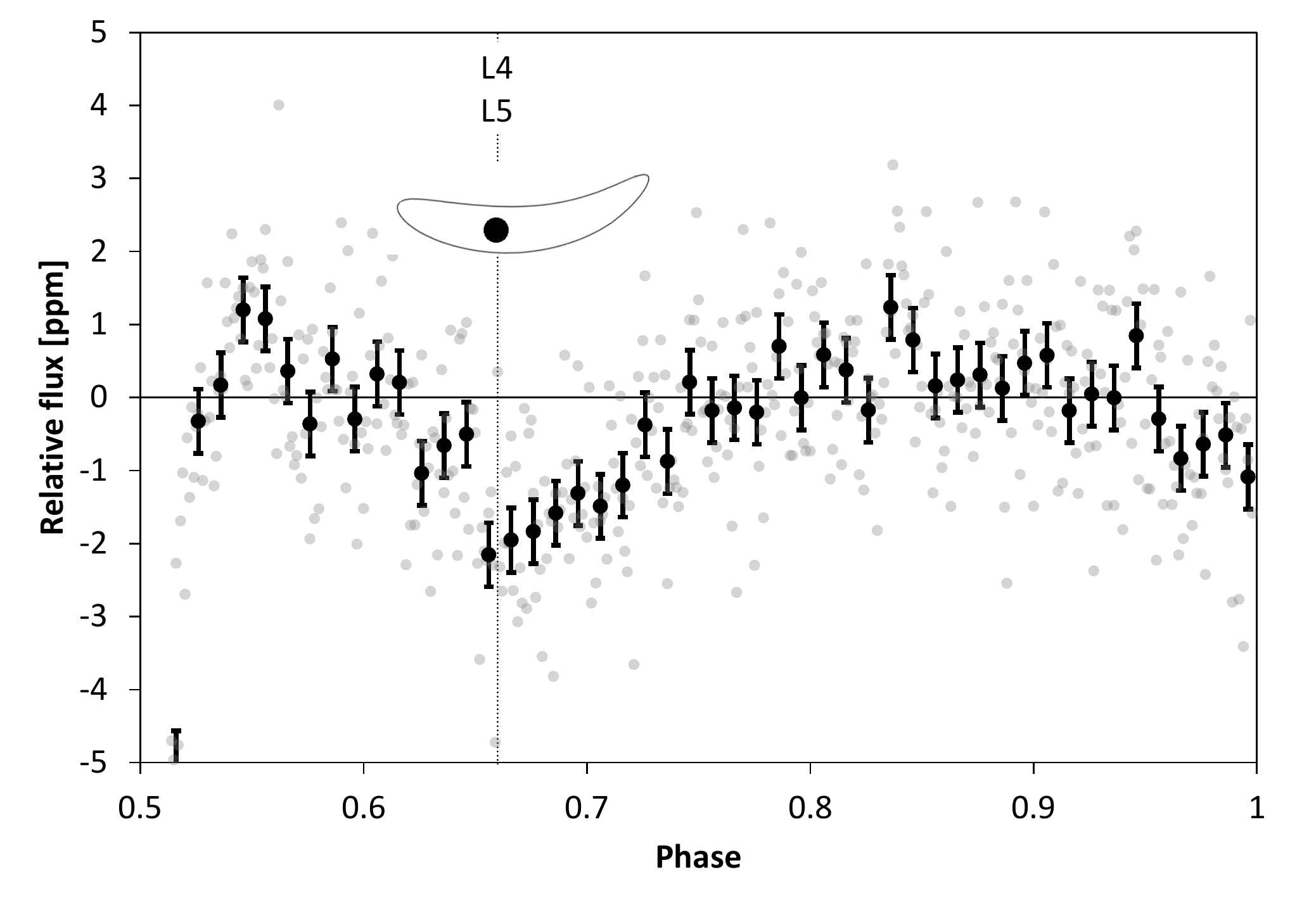}
\caption{\label{fig:l45}Sub-sample superstack in normal (left) and double symmetric (right) phase fold, with expected orbit size shown for reference. Note different vertical axes. Gray dots are 1,000 bins over phase space, black dots with error bars (right) are 100 bins for better visibility.\\}
\end{figure*}

\begin{figure}
\includegraphics[width=1.1\linewidth]{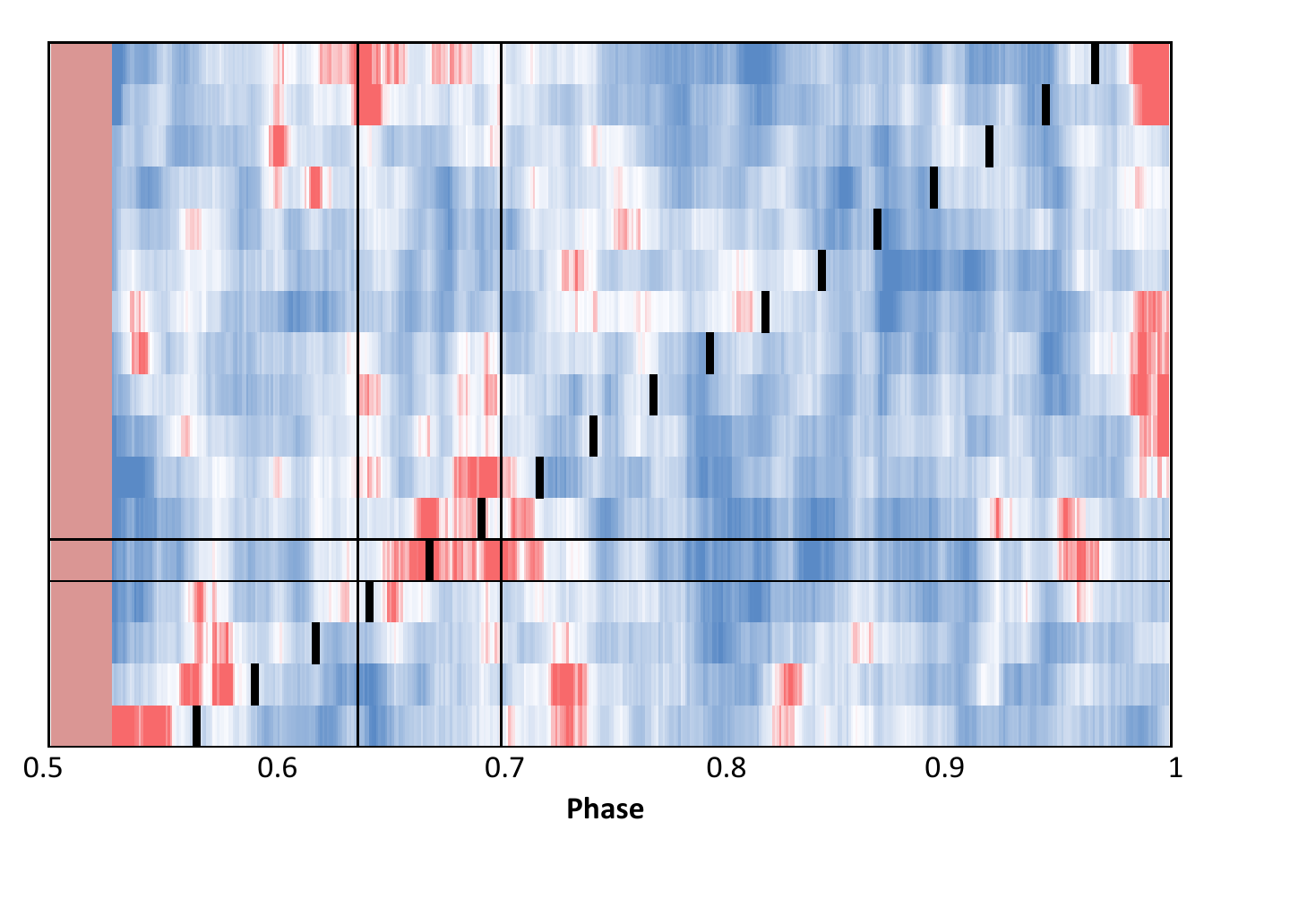}
\caption{\label{fig:sanity}Cross-check of sub-sample selection artifacts. In each line, we select those data that have a dip on one side of phase space, and plot their flux only for the other half of phase space. For example, in the first row, all data are shown that have a dip at phase 0.034 (at boxcar width 0.03). If an artifact were present, we would expect a corresponding dip (red color) at phase $1-0.34=0.966$, which is not seen. Also, we would expect such an artifact to occur in every line, centered at the black marked, which is not the case. Instead, we mainly see red dips occur at phase $\sim0.66$, where the Trojan transits are expected. A few columns (0.025 in phase time) around primary transits are excised as these values are $\sim1000\times$ deeper in flux, making them incompatible with useful color scalings.}
\end{figure}

\subsection{Cross-check for secondary eclipses}
\label{sub:secondaries}
As a useful cross-check for our data preparation method, we have searched for the average secondary eclipse. For simplicity, we have assumed only reflected light with an average albedo of 0.22 \citep{Sheets2014} and neglected differences in temperatures. As can be seen in Figure~\ref{fig:fold}, there is only a hint of a secondary feature at phase 1.0 = 0.0, which is measured to be $-0.31\pm0.21$ppm at a bin width of 0.01. Following our simplified assumptions, we can calculate the expected dip for this sample as $(R_P/a)^2$ per planet, giving an average of -0.88ppm for the sample. We explain the difference as caused by smearing from shifts in transit timing (from non-zero eccentricity) and different transit durations of each planet, which we did not compensate for.

We have also checked a different sample which is expected to yield a higher secondary dip: All planets with radii $>2R_{\oplus}$ on orbits $<0.3$au. This sub-sample is expected to yield an average secondary eclipse of -1.7ppm; our data analysis gives $-0.73\pm0.34$ in the same bin width. Again, we have to expect centering variations which reduce (broaden) the observed depth. It is however reassuring to see a dip at $>2\sigma$ significance. Finally, we have checked the few individual examples from \citet{Sheets2014} where the secondary eclipse is detected for individual planets (e.g. Kepler-10b); these dips are also present in our dataset. We conclude that there seems to be no obvious fault in our dataset, and that secondary eclipses are hard to detect for most of the \textit{Kepler} planets.

\subsection{Sub-sample analysis}
\label{sub:sample}
The full sample might be heavily diluted by a large number of systems with no, or relatively few, Trojans. We test this hypothesis by assuming that the flux in L4 and L5 is uncorrelated for any other cause than Trojan bodies. We know from our own solar system that the number of bodies in L4 and L5 is approximately equal. Then, we can examine a sub-sample of the superstack: We take all those planets that exhibit a negative flux at L4 (phase 0.33), and take their data of phase $0.5..1$ for further analysis. The same is done for L5 in the reverse logic. This gives us $1,251$ samples of negative flux at phase 0.33, and their light curves for the ``right part'', i.e. flux phase $0.5..1$. We also find $1,266$ samples with a dip at phase 0.66, and take their part of the light curves from phase $0..0.5$. Afterwards, we stitch these halves together, and obtain 1,940 lightcurves (some have dips in both halves). The result of this sub-sample is shown in Figure~\ref{fig:l45} and exhibits a clear dip at both L4 and L5, with a maximum depth of 2ppm (970km radius equivalent). It is re-assuring to see that this dip is not uniform; as can be seen in the double-phase fold (right part of this figure), its shape is elongated away from planetary transit, as is expected for distributions from horseshoe and tadpole orbits. 

An alternative interpretation of the dips in figure~\ref{fig:l45} (left) would be numerical fluctuations, caused by autocorrelation.  Indeed, a Durbin-Watson test returns clear autocorrelation ($p=0.01$) if the complete dataset of 1,000 bins is used. However when we excise the phase times with signals (around 0, 0.33, 0.5, 0.66 and 1) and treat only the remaining data, then autocorrelation is insignificant even at the 10\% level.

We have also cross-checked whether this dip is introduced by some symmetry artifact. When selecting any other phase-folded time, e.g. flux $<0$ at phase 0.2, no equivalent dip on the ``other side'' of the orbit, i.e. at phase 0.8, can be reproduced. Figure~\ref{fig:sanity} shows this: If an artifact were present, we would expect a dip centered at each red mark, which is not present. Clearly, we cannot produce a similar dip at any phase time with this symmetry argument; it only works at phase times 0.33 (L4) and 0.66 (L5). We caution, however, that the S/N of the total signal is low, as will be explained in the following section. Splitting such a weak signal into different views can therefore only create weak indicators of its validity.

\subsection{Significance of the result}

To measure the significance of these dips, we take the signal-to-noise ratio for transits \citep{Jenkins2002,Rowe2014}, which compares the depth of the transit mode compared to the out-of-transit noise:

\begin{equation}
\label{eq:sn}
{\rm S/N} = \sqrt{N_{T}}\frac{T_{dep}}{\sigma_{OT}}
\end{equation}

with $N_{T}$ as the number of transit observations, $T_{dep}$ the transit depth and $\sigma_{OT}$ the standard deviation of out-of-transit observations. For the Lagrangian signals, we find the L4 and L5 dip at S/N$\sim$6.7 each, and a combined S/N=9.3. It has been argued by \citet{Fressin2013} that the detection of transits becomes unreliable for a S/N $\lesssim$ 10, so that this signal can only qualify as a tentative detection.

\begin{figure}
\includegraphics[width=\linewidth]{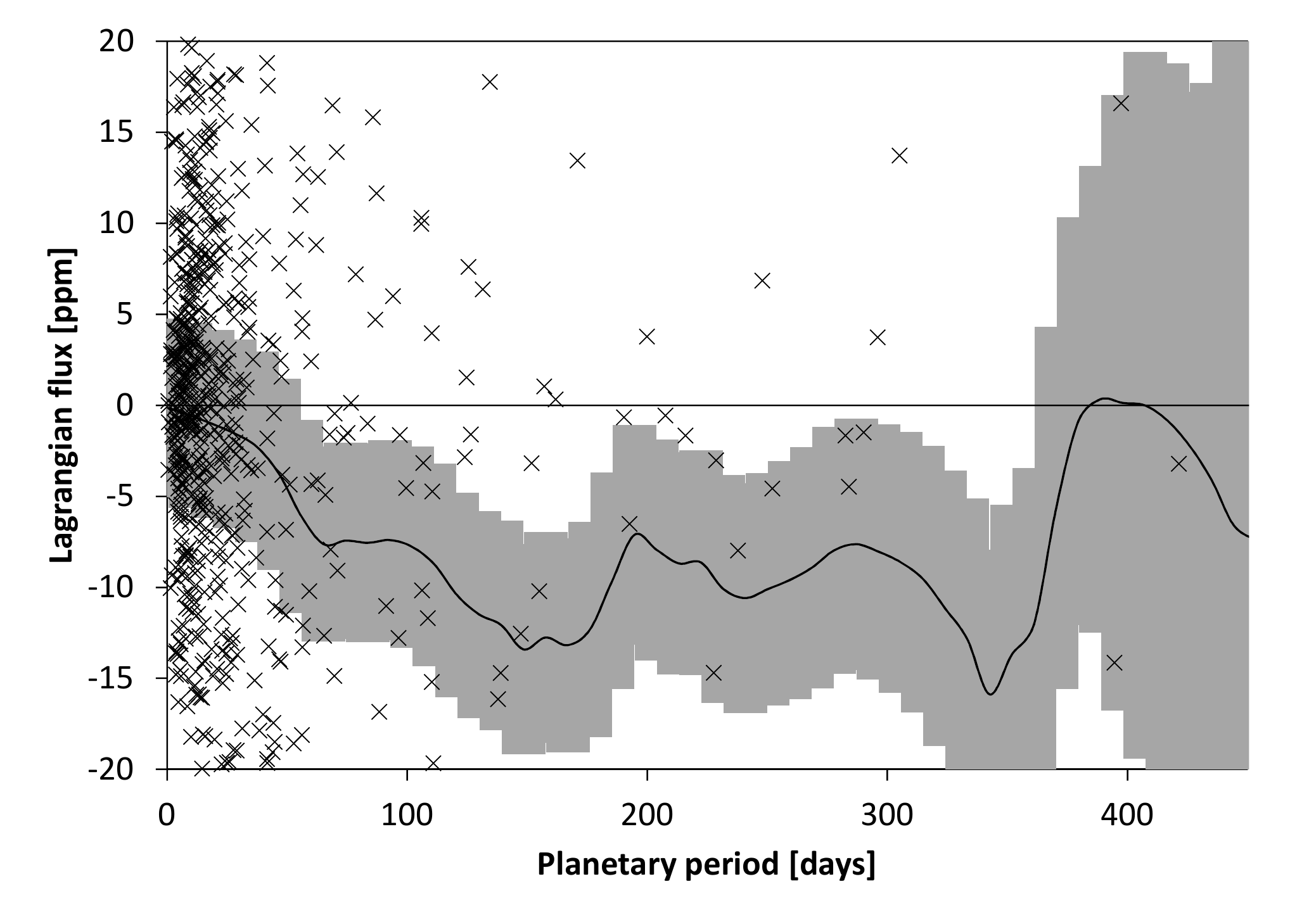}
\caption{\label{fig:breakdown}Density estimate for the Trojan-like signal versus planetary period. The shaded area is the local $2\sigma$ confidence band. Short periods $<50$ days are consistent with zero trojan signal, but a trojan-like signal is detected for periods between 60 and 350 days. Uncertainty increases for longer periods due to a lack of data. See text for discussion.}
\end{figure}

\subsection{Sub-sample properties}
We have compared the properties of the 1,940 planets in our sub-sample to the total \textit{Kepler} sample. We use a nonparametric density estimation with a local polynomial regression to include local confidence bands (e.g., \citet{Ruppert2003, Takezawa2006}).

We find a correlation of the Trojan-like dips to the period of the host star: At $p>20$ days, the probability density moves towards more pronounced Trojan-like signal, but the effect becomes only formally significant for $60<p<350$ days. This might either reflect a stability dependence of Trojan bodies to their semi-major axis, or a formation bias, or a mix of both. Due to radiation effects (such as the Yarkovsky effect \citep{Bottke2006}, which can cause small objects to undergo orbital changes), we can expect few (if any) close-in ($p<10$ days) small asteroids. For long periods, $p>350$ days, the sample size is too small for a significant result.

We have tried the same density estimates for several other measures, but all are formally insignificant. At first glance, for example, one might assume that the impact parameter of the planetary transit could positively affect trojan detectability: For more central planetary transits, one might assume a sky-coplanar trojan to also (more likely) transit centrally, making detection easier. This is, however, likely overpowered by the inclination scatter of possible trojan ``swarms''. If we take the inclination scatter of Jupiter's trojans as a proxy, then the projected size of a trojan cloud in exoplanetary systems would be several times larger than the projected size of their star. Given our data quality, it is not surprising to find no significant correlation with respect to the impact parameter.

The same argument can be made for a correlation to the stellar radius. One might hypothesize that the total trojan mass correlates to the total mass of the circumstellar disk, which itself might be correlated with the mass of the star. More massive stars are known to host more massive planets \citep{Johnson2010}, so that Trojan bodies might also be more massive (i.e., larger at a given density). The scale for such a correlation, however, is expected to be of order $R_{trojan} \sim M_{trojan}^{1/3}$. The majority of Kepler planets and candidates are found for star between $0.5 R_*$ and $2 R_*$, a range which, in combination with the limited data quality, does not allow for the detection of a significant correlation of trojan occurrence to the stellar radius.

Finally, we also tried correlations to the planetary radius, multiplicity, metallicity of the host star, and eccentricity; all of which gave null results.

\section{Discussion}
\label{sec:discussion}
While the data quality from \textit{Kepler} is the best we have, it is only barely sufficient to search for Trojan bodies. Still, we believe that the methods outlined in this paper will be valuable in the future, when more and better data come available. The \textit{PLATO 2.0} mission \citep{Rauer2014} will deliver photometry for 500 bn stars in the years after 2024, and up to $3\times$ better photometric precision. With such a dataset, the analysis performed here should be repeated, and should yield highly significant results for every breakdown. Also, a few single large Trojans might also be expected, if the vast dataset \citep{Hippke2015b} can be mined sufficiently. 

We have explored the potential of \textit{PLATO 2.0} using lower-limit estimates from \citet{Rauer2014} for its scientific return. Then, $\sim10\times$ more lightcurves will be available, when compared to \textit{Kepler}, for a duration of 6 (instead of 3) years. For simplicity, we neglect the better instrumental noise properties. This gives $1/\sqrt{2\times10}$ of noise improvement per bin (compared to the current data), resulting in $\sim0.3$ppm of noise in each of the 1,000 bins. Consequently, from a superstack without any selections, we can expect equal or better signal-to-noise properties than in the heavily selected and biased \textit{Kepler} sample. More precisely, the full \textit{PLATO 2.0} sample is expected to yield a signal-to-noise as shown in Figure~\ref{fig:l45}, without any of our discerning selection choices.

\begin{figure}
\includegraphics[width=1.1\linewidth]{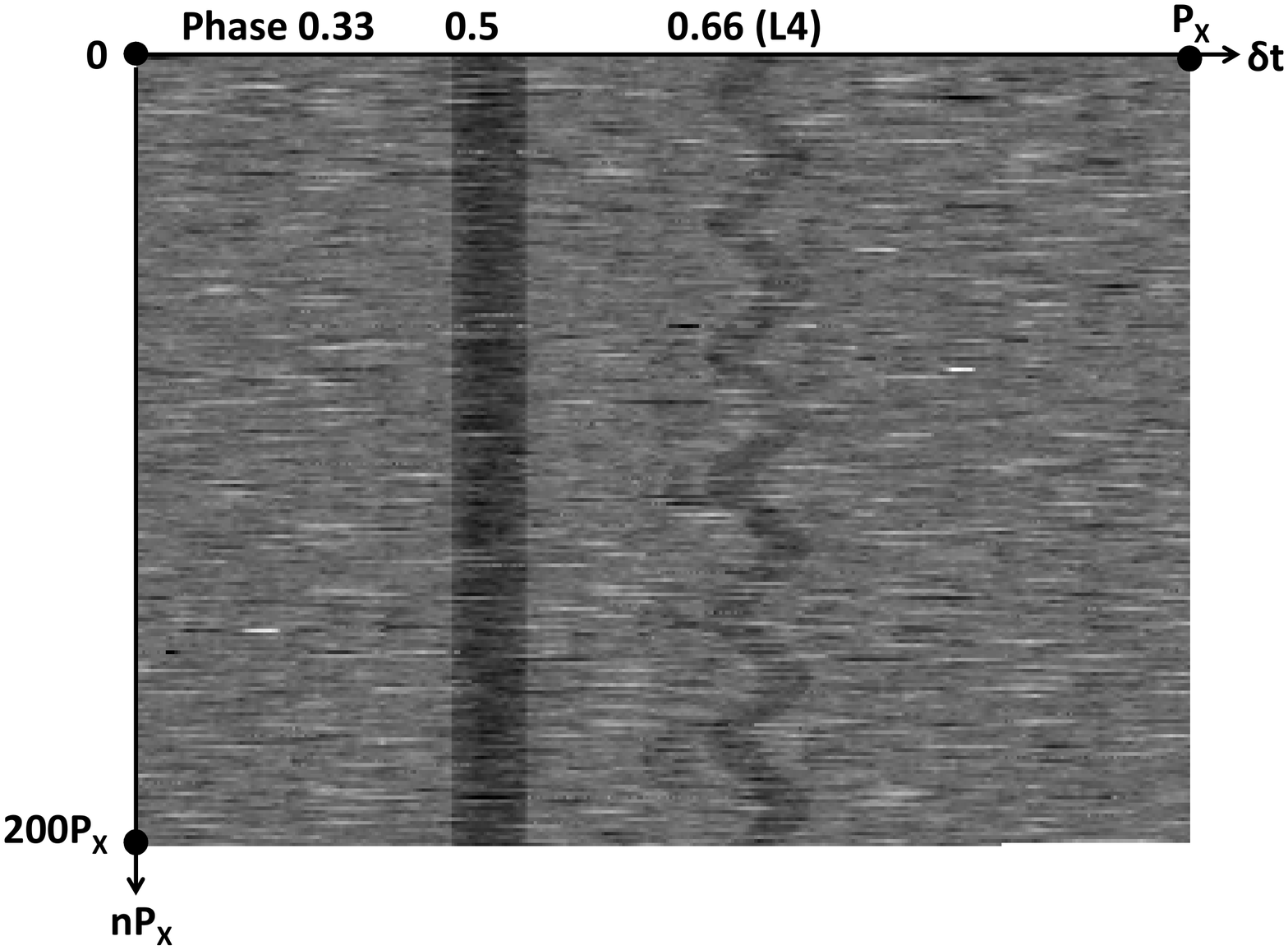}
\caption{\label{fig:plato}Riverplot of Trojan injection into 6 years of real solar data, assuming \textit{PLATO 2.0} instrumental performance. A synthetic 10-day period Hot Jupiter is visible at phase 0.5; the injected 0.8R$_{\oplus}$ Trojan is clearly visible in its horseshoe-orbit at L4.}
\end{figure}

Furthermore, we can expect to make clear detections of large individual Trojans with \textit{PLATO 2.0}, if such bodies exist with transiting areas $>0.5R_{\oplus}$. To show this, we have created a series of injections. Our process is similar to the one described in \citet{Hippke2015b}. In short, we take real solar data from \textit{VIRGO/DIARAD} \citep{Frohlich1997} and add instrumental noise from an end-to-end \textit{PLATO~2.0} simulator \citep{Zima2006,MarcosArenal2014}. Into these data, we inject synthetic Trojan lightcurves, assumed to orbit in horseshoe-orbits around L4/L5 with semi-major axes of $\sim0.02$ in phase-time \citep{Janson2013}. To explore the parameter space of recoverable signals, we varied the transiting Trojan area (radius), the stellar radius, and the planetary period. We show an exemplary riverplot in Figure \ref{fig:plato} for a Sun-like star, orbited by a 10-day Hot Jupiter and a 0.8R$_{\oplus}$ Trojan in L4. Such a configuration is easily detected in visual examination, but might escape standard transit detection algorithms due to the large transit timing variations. We find that for Sun-like stars, transiting areas of $>0.65$R$_{\oplus}$ (Mars-size) are easily detected visually; the equivalent limit for a 0.5R$_{\odot}$ M-Dwarf is $\sim$0.5R$_{\oplus}$. Instead of a visual search, algorithms might be used, as explained by \citet{Janson2013}. It is however unclear how efficient these can be; this would have to be determined with a series of injections and (blind) retrievals.

An interesting finding from our own injections is that potential Trojans at shorter period planets are much more easily identified due to their higher number of transits. Our example uses a 10-day planet; this is on the lower end of our hypothesized limit of planets having Trojans, as explained in section \ref{sub:sample}. For longer period planets, e.g. a 100-day planet, the number of rows (transits) in Figure~\ref{fig:plato} would be 20 (instead of 200) for 6 years of data. This reduces chances of detection, highlighting the benefits of long-term campaigns.

\section{Conclusion}
\label{sec:conclusion}
With the given dataset, we only find a significant Trojan-like signal when applying the ``left-right'' method, selecting only L5 data for those candidates that seem to exhibit a L4 dip (and vice versa). While we tested this method to be robust against a symmetrical bias, it also implies that the main sample must be heavily diluted with a large number of systems with no (transiting) Trojans. If the method is valid, then the breakdowns of this sub-sample indicate that Trojans are more prominently found for longer ($>$60~days) periods. These cautious, and preliminary findings might inspire theorists to advance planetary formation theory in that direction; these theories can then be validated with the upcoming data from the \textit{PLATO 2.0} spacecraft.

\end{document}